\documentclass[
reprint,
 amsmath,amssymb,
 aps,
floatfix,
]{revtex4-2}
\usepackage{graphicx}
\usepackage{dcolumn}
\usepackage{bm}
\usepackage{float}
\usepackage{mleftright}
\usepackage{mathtools}
\usepackage{mleftright}
\usepackage{physics}
\usepackage{xcolor}

\begin{document}

\preprint{APS/123-QED}
\title{Symmetry breaking in time-dependent billiards}
\author{$^1$Anne Kétri P. da Fonseca, $^1$Edson D. Leonel}
\affiliation{$^1$Departamento de Física, UNESP - Universidade Estadual Paulista, Avenida 24A, 1515, Bela Vista, Rio Claro, 13506-900, São Paulo, Brazil}
\date{\today}
\begin{abstract}
We investigate symmetry breaking in a time-dependent billiard that undergoes a continuous phase transition when dissipation is introduced. The system presents unlimited velocity, and thus energy growth for the conservative dynamics. When inelastic collisions are introduced between the particle and the boundary, the velocity reaches a plateau after the crossover iteration. The system presents the expected behavior for this type of transition, including scale invariance, critical exponents related by scaling laws, and an order parameter approaching zero in the crossover iteration. We analyze the velocity spectrum and its averages for dissipative and conservative dynamics. The transition point in velocity behavior caused by the physical limit of the boundary velocity and by the introduced dissipation coincides with the crossover interaction obtained from the $V_{rms}$ curves. Additionally, we examine the velocity distributions, which lose their symmetry once the particle's velocity approaches the lower limit imposed by the boundary's motion and the system's control parameters. This distribution is also characterized analytically by an expression $P(V,n)$, which attains a stationary state, with a well-defined upper bound, only in the dissipative case.
 
  \end{abstract}                            
\maketitle

\section{Introduction}
Phase transitions have been extensively studied for more than a century, with applications spanning various physical systems. From the development of thermodynamics in the 19th century to the statistical mechanics of the 20th century, the concept of phase transitions has evolved to encompass increasingly complex phenomena \cite{phase1}. A significant milestone was the identification of phase transitions that involve the breaking of continuous symmetries, most notably the Bose-Einstein condensation observed in ideal Bose gases as the temperature is lowered at a fixed positive density, and subsequently focus on systems with discrete symmetries, such as the Ising model \cite{principles}. Currently this concept has found several relevant applications in quantum many-body physics \cite{manybody1,manybody2}, ultra-cold atomic systems \cite{Sadhasivam2024}, high-energy physics \cite{Aarts2023}, information theory \cite{info1,info2} and, as in this case, dynamical systems \cite{leonel2015dynamical,oliveira2013some,sethna2021statistical}.

Different phases of a system can usually be assigned to different symmetries: the sudden rearrangement of a crystal lattice changes the state of the body discontinuously between different phases \cite{landau}. These changes in symmetry may also happen continuously, in second-order, also called continuous phase transitions, named after the corresponding order of the free energy derivative that breaks continuity during the transition \cite{ehrenfest1933phase}. We can refer quantitatively to this change in symmetry, defining the quantity called "order parameter", which continuously approaches zero as the system reaches the transition point \cite{landau}. 

These continuous phase transitions are also characterized by their scale invariance, with critical exponents related by scaling laws, and which do not depend on the microscopic details of the system, but only on its dimensionality and symmetry, a phenomenon known as universality \cite{reif2009fundamentals}. Such examples of these transitions are the Ising model, where the magnetization vanishes as the temperature is raised until it's critical value \cite{sethna2021statistical}, and the transition from integrability to non-integrability for the standard dissipative mapping\cite{leonel2020characterization}, where it changes from regular to chaotic behavior in the phase space. 

As mentioned, a key step to characterize such phase transitions is to identify the symmetry broken at the critical point, which marks a qualitative change in the system's macroscopic behavior \cite{sethna2021statistical}. In this work, we characterize such a symmetry break for the phase transition observed in time-dependent billiards from the conservative to the dissipative case. 

A billiard is a dynamical system composed of a particle or a set of non-interacting particles suffering specular collisions with a rigid boundary that confines them \cite{chernov2006chaotic}. The shape of the boundary plays a fundamental role in determining the system's behavior. Depending on its geometry, the dynamics can range from fully integrable, as in the case of the circular billiard \cite{berry1981regularity,bunimovich2005open}, to fully chaotic, as exemplified by the Sinai billiard and the Bunimovich stadium \cite{dettmann2009survival,lozej2018aspects}, or exhibit a mixed phase space structure, as seen in the oval billiard \cite{lopac2002chaotic}. Billiards can be framed by Hamiltonians such as  $H(x,p,t) = \frac{p^2}{2m} + V(x,t),$ where \( V(x,t) = V_0(x) + V_1(x,t) \). The component \( V_1(x,t) \) is associated with the time-dependency introduced in the boundary, leading to non-integrability \cite{lichtenberg2013regular}.

Although seemingly very simple, these systems offer a wide array of potential applications across diverse fields, including optics and photonics \cite{PhysRevLett.84.867,Wang2021}, plasma physics \cite{Berglund1996}, electronic transport \cite{Electronictransport}, complex networks \cite{EXNER2006445}, and condensed matter physics \cite{science.1144359}. Therefore, characterizing its symmetry breaking is extremely important for better understanding the mechanisms behind these systems' phase transitions.

In this work, we investigate the symmetry breaking in an oval-shaped billiard as it goes from unbounded to bounded diffusion as dissipation is introduced. The boundary is given by
$R_b(\theta,t) = 1 + \epsilon[1 + \eta \cos(t)] \cos(p\theta)$, where \( \epsilon \) controls the integrability of the system and \( \eta \) gives the amplitude of the temporal dependence. A discrete four-dimensional mapping describes the system's dynamics \( T \). The system is integrable system for \( \epsilon = 0 \), and presents mixed dynamics for \( \epsilon \neq 0 \) \cite{chernov2006chaotic}. The Loskutov-Ryabov-Akinshin (LRA) conjecture determines that the existence of chaos in the phase space is a sufficient, not necessary, condition for the occurrence of an unlimited energy growth when a time perturbation of the boundary is introduced \cite{loskutov1999mechanism,loskutov2000properties}. This energy growth is also known as Fermi acceleration (FA). This phenomenon can be suppressed by introducing dissipation, in this case in the form of inelasticity in the collisions characterized by a restitution coefficient $\gamma$.

In the conservative case, the particle velocities exhibit FA due to diffusive behavior, while the dissipative case leads the system to a stationary state. We show that the onset of this transition is governed by the boundary velocity's physical limit and the dissipation's strength, as revealed by the $V_{rms}$ and $\overline{V}$ curves. The velocity distribution becomes asymmetric near its lower bound, determined by the system's control parameters, and reaches a stationary profile only in the dissipative regime. This behavior is also recovered by the analytical distribution $P(V,n)$ near this transition.

This paper is organized as follows: Section \ref{xsec2} describes the model for the time-dependent oval billiard and discusses its velocities behavior, along with scaling properties Sections \ref{xsec3} and \ref{xsec4} present the behavior of such velocities for the conservative and dissipative case, respectively, as well as the velocities distributions for each instances, comparing with the established results for this system. Section \ref{xsec5} presents the final discussions and conclusions regarding the symmetry breaking.

\section{The model and the mapping}
\label{xsec2}
As mentioned, the main focus of this work is the oval billiard. The boundary is given, in polar coordinates, by \( R_b(\theta,t) = 1 + \epsilon[1 + \eta \cos(t)] \cos(p\theta) \), where \( \theta \) is the polar angle, \( \epsilon \) is a parameter controlling the circle deformation, \( \eta \) gives the amplitude of the time perturbation, and \( p \) deforms the boundary. We limit our analysis to integer values of $p$, as non-integer values would lead to the existence of holes in the boundary, allowing the escape of particles and going beyond the scope of this work. $\epsilon=0$ retrieves the circular billiard, which we know to be fully integrable, while \( \epsilon \neq 0 \) leads to a mixed phase space, containing both chaos, periodic islands, and invariant spanning curves \cite{lopac2002chaotic}. The further increase of $\epsilon$ leads to the vanishing of these regular structures, being fully chaotic for values above $ \epsilon_c = \frac{1}{1 + p^2} $ \cite{oliveira2010suppressing}. The time-dependent boundary \( \eta \neq 0 \), as defined by the LRA conjecture \cite{loskutov2000properties}, leads to unbounded diffusion caused by chaotic dynamics for the static regime. 

We can write a four-dimensional nonlinear mapping $T(\theta_n, \alpha_n, V_n, t_n)$ welding the dynamics of the impact \( n \) with \( n+1 \). $\theta$, as mentioned, represents the polar angle relative to the origin of the coordinate system, while $\alpha$ corresponds to the angle between the trajectory and the vector tangent to the boundary. $V$ and $t$ represent, respectively, the velocity of the particle and the instant of impact. The coordinates for the particle are given by $X(\theta_n)=R(\theta_n,t_n)\cos(\theta_n)$ and $Y(\theta_n)=R(\theta_n,t_n)\sin(\theta_n)$. The velocity is then written as:
\begin{equation}
    \vec{V_n}= |\vec{V_n}|[\cos(\phi_n+\alpha_n)\hat{i} + \sin(\phi_n + \alpha_n)\hat{j}],
\end{equation}
where $\phi_n= \arctan\left[Y' (\theta_n,t_n)/{X' (\theta_n,t_n)}\right]$ is an auxiliary angle, with primes denoting partial derivatives over $\theta_n$. The specular reflection law must also be obeyed, for the non-inertial referential frame of the moving boundary, this is written as:
\begin{eqnarray}
    \vec{V'}_{n+1} \cdot \vec{T}_{n+1} =  \vec{V'}_{n} \cdot \vec{T}_{n+1} \\
    \vec{V'}_{n+1} \cdot \vec{N}_{n+1} =  \vec{V'}_{n} \cdot \vec{N}_{n+1}
\end{eqnarray}
where $\vec{T}$ and $\vec{N}$ indicate the unit vectors for the tangential and normal components, respectively, written as $\vec{T}_{n+1}=\cos(\phi_{n+1})\hat{i}+\sin(\phi_{n+1})\hat{j}$ and $\vec{N}_{n+1}=-\sin(\phi_{n+1})\hat{i}+\cos(\phi_{n+1})\hat{j}$. With these equations, the velocity at the $(n+1)^{th}$ collision is finally written as:
\begin{equation}
\vert \vec{V}_{n+1}\vert = \sqrt{(\vec{V}_{n+1}\cdot \vec{T}_{n+1})^2 + (\vec{V}_{n+1}\cdot \vec{N}_{n+1})^2}.
\end{equation}
The trajectory of the particle is given by:
\begin{eqnarray}
X(t) &=& X(\theta_n ,t_n) + \vert \vec{V_n}\vert \cos(\alpha_n + \phi_n)(t-t_n), \\
Y(t) &
=& Y (\theta_n ,t_n) +  \vert \vec{V_n}\vert \sin(\alpha_n + \phi_n)(t-t_n).
\end{eqnarray}
The position of the particle can then be obtained by $R(t) =\sqrt{X^2 (t) + Y^2 (t)}$. The point of collision is found when $R = R_b$. After the collision, the updated angle $\alpha_{n+1}$ is given by $\alpha_{n+1}=\arctan\left[\vec{V}_{n+1}\cdot \vec{N}_{n+1}/{ \vec{V}_{n+1}\cdot \vec{T}_{n+1}}\right]$.

To analyze such velocities, it is convenient to recur to two new quantities: $\overline{V}$ and $V_{rms}$. For $M$ different initial conditions and $n$ collisions of the particle with the boundary, they are written as:
\begin{equation}
V_{rms} = \sqrt{\overline{V^2}(n)}=\sqrt{\frac{1}{M} \sum_{i=1}^{M} \frac{1}{n} \sum_{j=1}^{n} V_{i,j}^2}.
\end{equation}
Here \( V_{i,j} \) represents the velocities for each trajectory \( i \) and for all \( j \) different orbits.

For the conservative case, as shown in Fig.1 a), the behavior of \( V_{rms} \) can be summarized as follows: (i)  \( V_{rms} \propto n^\beta \) with \( \beta \approx 0.5 \) for small values of (\( V_0 \)); (ii) A plateau \( \overline{V}_{plat} \propto V_0^\zeta \) is observed for \( n \ll n_x \), where \( \zeta \approx 1 \); (iii) the crossover iteration between (i) and (ii) is given by \( n_x \propto V_0^z \), with \( z \approx 2 \). The scale invariance observed in the system allows us to, using a homogeneous and generalized function, write a scaling law relating the three exponents \( z = \zeta / \beta \).  The overlap of the curves into a single universal one can be done using the scaling transformations $V_{rms} \rightarrow V_{rms}/V_0^\zeta$ and $n\rightarrow n/V_0^z$. The universal plot is given in Fig.1 b).
\begin{figure}[h]
\centering
\includegraphics[width=1\linewidth]{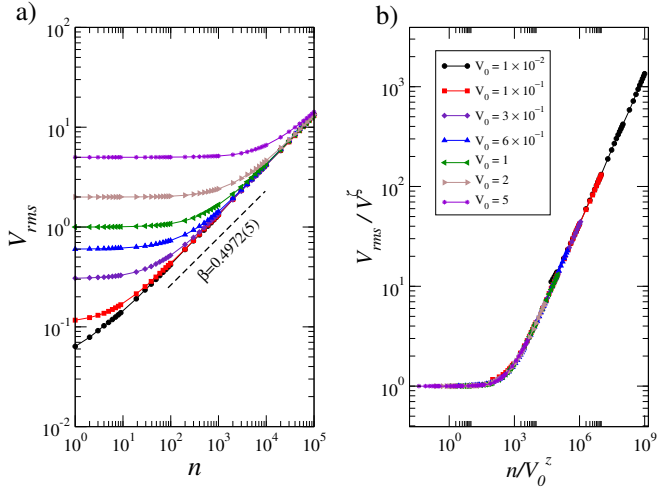}
\caption{(Color online) (a) $V_{rms}$ $vs.$ $n$ constructed from the analytical expression for and $\epsilon=0.08$, $p=3$, $\eta=0.5$ and different values of $V_0$. (b) Overlap of the curves in \textit{(a)} into a single universal curve using the scaling transformations $V_{rms} \rightarrow V_{rms}/V_0^\zeta$ and $n\rightarrow n/V_0^z$.} 
\end{figure}
As mentioned, the conservative case leads to unlimited velocity and, thus, energy growth, known as FA. Our chosen method to suppress this behavior is to consider the collisions of the particles with the boundary to be inelastic. The dissipation is introduced by a restitution coefficient $\gamma<1$ in the standard velocity component. For $\gamma=1$, the collisions are elastic, returning to the conservative case. The updated reflection law is:
\begin{equation}
\vec{V}_{n+1} \cdot \vec{N}_{n+1} = -\gamma \vec{V}_{n} \cdot \vec{N}_{n+1} + (1 + \gamma) \vec{V_b}[t_{n+1} + Z(n)] \cdot \vec{N}_{n+1}
\end{equation}
Figure 2 a) shows that fractional energy loss does indeed suppresses the FA. We can now write a new set scaling hypotheses as: (i) \( V_{rms} \propto [(\eta \epsilon)^2 n]^\beta \) for \( n \ll n_x \), with \( \beta \approx 0.5 \); (ii) the velocity reaches a plateau described by \( \overline{V}_{sat} \propto (1 - \gamma)^{\zeta_1} (\eta \epsilon)^{\zeta_2} \), for \( n \gg n_x \), where \( \zeta_1 \approx -0.5 \) and \( \zeta_2 \approx 1 \); (iii) the crossover iteration is given by \( n_x \propto (1 - \gamma)^{z_1} (\eta \epsilon)^{z_2} \), with \( z_1 \approx -1 \) and \( z_2 \approx 0 \).  The two scaling laws for this case are $z_1=\frac{{\zeta_1}}{\beta}$ and $z_2=\frac{\zeta_2}{\beta}-2$. Again overlap of the curves into a single universal one can be done using scaling transformations: $V_{rms} \rightarrow V_{rms}/(1-\gamma)^{\zeta_1}(\epsilon\eta)^{\zeta_2}$ and $n\rightarrow n/(1-\gamma)^{z_1}(\eta\epsilon)^{z_2}$. The curves overlap into a single universal plot, as given in Fig. 2b.
\begin{figure}[h]
\centering
\includegraphics[width=1\linewidth]{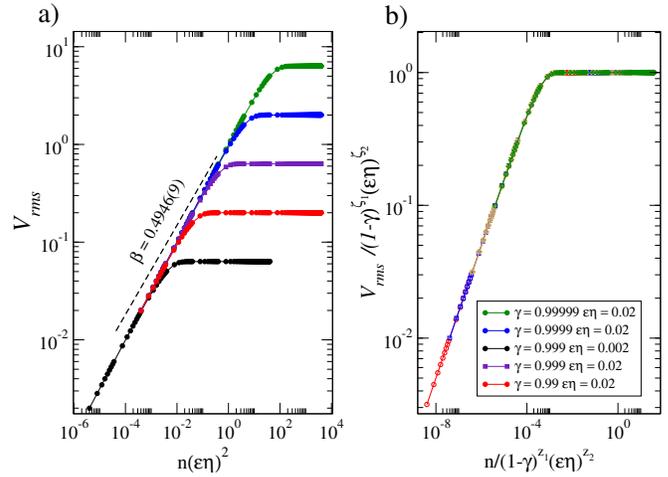}
\caption{ (Color online) (a) $V_{rms}$ $vs.$ $n(\epsilon \eta)^2$ constructed from the analytical expression for initial velocity $V_0 = 10^{-5}$ for different values of $\gamma$ and $\eta \epsilon$. (b) Overlap of the curves in \textit{(a)} into a single universal curve using the scaling transformations $V_{rms} \rightarrow V_{rms}/(1-\gamma)^{\zeta_1}(\epsilon\eta)^{\zeta_2}$ and $n\rightarrow n/(1-\gamma)^{z_1}(\epsilon\eta)^{z_2}$.} 
\end{figure}
Our main goal is to investigate the broken symmetry between the two behaviors, presenting the velocity distributions, the averages, and their relationship with $V_{rms}$, as discussed in the following sections.

\section{Conservative case}
\label{xsec3}
To better characterize the symmetry breaking, we examine both the root-mean-square velocity ($V_{\text{rms}}$) and the complete velocity distribution Fig. 3 shows, for the first 100 collisions, the trajectories of 10 different initial conditions $(\theta_n,\alpha_n)$ randomly chosen from [$0$,$2\pi$] for $V_0=1$, $\epsilon=0.08$, $\eta=0.5$ and $\gamma=1$. The maximum and minimum values reached from these trajectories are saved, leading to the purple and turquoise lines, named $V_{max}$ and $V_{min}$, respectively. The $V_{rms}$ curve is green, still in the plateau region, as observed in Fig. 1 . The average values of the velocity for the regions above and below $V_{rms}$ are shown in the blue and red curves, respectively. We refer to this quantities as $\overline{V}_{max}$ and $\overline{V}_{min}$.

In agreement with the plateau observed for $V_{rms}$, the distribution of velocities of each trajectory is distributed symmetrically around the initial value $V_0$. The curves of $V_{max}$, $V_{min}$, $\overline{V}_{max}$ and $\overline{V}_{min}$ also present such symmetry. 
\begin{figure}[h]
    \centering
    \includegraphics[width=1\linewidth]{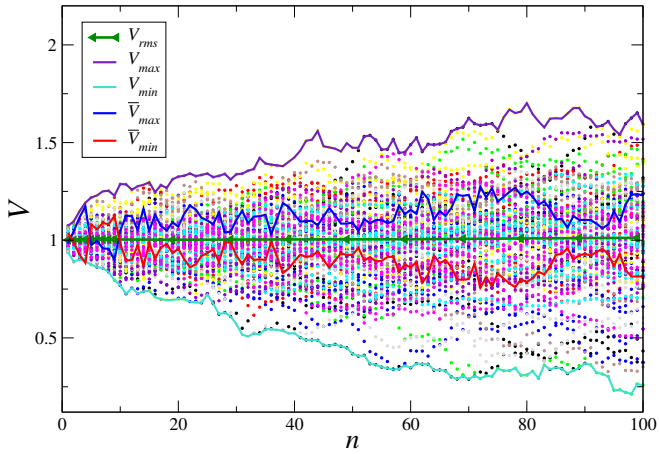}
    \caption{(Color online) $V_{rms}$, $V_{max}$, $V_{min}$, $\overline{V}_{max}$ and $\overline{V}_{min}$ vs. $n$ for $V_0=1$, $\epsilon=0.08$, $\eta=0.5$ and $\gamma=1$. The circles represent different trajectories traversed by $10$ initial conditions for each color.}
    \label{fig:enter-label}
\end{figure}
Fig.4 presents the evolution of the five curves $V_{rms}$, $V_{max}$, $V_{min}$, $\overline{V}_{max}$ and $\overline{V}_{min}$ for $V_0=0.6$, $\epsilon=0.08$, $\eta=0.5$, $p=3$, $\gamma=1$ and $n=10^4$ collisions. The symmetric behavior between the curves $\overline{V}_{max}$ and $\overline{V}_{min}$ is broken after approximately 200 collisions, as highlighted in panel b). As observed in Fig. 5, increasing the value of $n$ leads to a gradual flattening of the curve, eventually reaching a lower limit for the particle velocities. This limit is given by the velocity of the moving boundary itself, given by \cite{leonel2016thermodynamics}
\begin{align}
    \vec{V_b}(t_{n+1})=  \mleft.\dfrac{\dd{R(t)}}{\dd{t}}\mright\vert_{t_{n+1}} [cos(\theta_{n+1})\hat{i} + sen(\theta_{n+1})\hat{j}] \propto \eta\epsilon
\end{align}
Thus, once the particles reach the lower limit, their value can only increase, leading to the observed symmetry breaking, where the distribution ceases to be Gaussian, as shown in blue in Fig. 5. This behavior is also observed in Fig. 4 for the curves $V_{max}$ and $V_{min}$, with the first growing unboundedly and the second oscillating around the lower limit in the collision zone $V_b\propto\pm \eta\epsilon$. Together, these behaviors lead to the power-law growth of the $V_{rms}$ curve, in blue. The presence of a lower limit combined with the absence of an upper limit for such a system also leads, for a sufficiently large number of collisions, to the phenomenon known as superdiffusion, where the coefficient $\beta$ becomes equal to $1$, as characterized by \cite{hansen}. The crossover iteration, $n_x$, then marks the point at which the average of the velocities reaches the lower limit. The blue distribution, for $n=100$, as shown in Figure 5, corroborates this analysis, with the appearance of a "tail" to the right of $V_0$. The same behavior is also corroborated by the probability density function $P(V)$ for the same system, shown in green in Fig. 6. This distribution has as boundary conditions \( P(V) |_{V \rightarrow 0} = P(V) |_{V \rightarrow \infty} = 0 \) and \( P(V)_{n=0} = \delta(V - V_0) \), ensuring that all particles start with the same initial velocity, distributed across \( M \) different initial conditions, uniformly distributed over \( \alpha \), \( \theta \), and \( t \). For a sufficiently small number of collisions, all velocities are centered around the initial velocity, according to the delta function. As the number of collisions increases, this distribution gradually flattens, remaining symmetric for $n=100$.
\begin{figure}[H]. 
    \centering
    \includegraphics[width=1\linewidth]{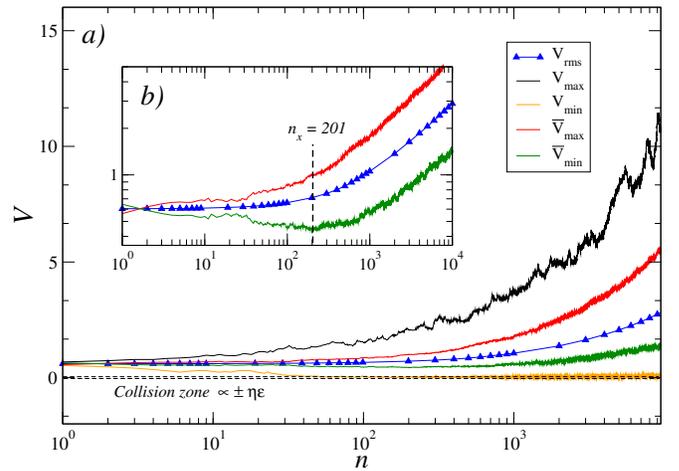}
    \caption{(Color online) a) $V_{rms}$, $V_{max}$, $V_{min}$, $\overline{V}_{max}$ and $\overline{V}_{min}$ vs. $n$ for $V_0=0.6$, $\epsilon=0.08$, $\eta=0.5$ and $\gamma=1$. Panel (b) shows a magnified view of the crossover iteration $n_x$ obtained from the green curve.}
    \label{fig:enter-label}
\end{figure}
\begin{figure}[H]
    \centering
    \includegraphics[width=.8\linewidth]{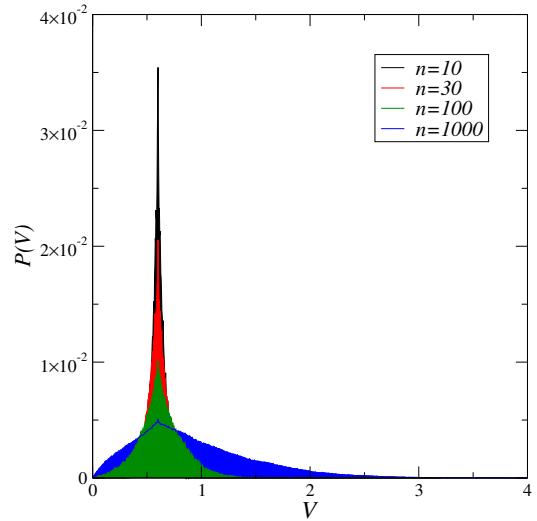}
    \caption{(Color online) Probability density for a set of 3500 particles after $n=10$, $30$, $100$ and $1000$ collisions. The parameters used were $V_0=0.6$, $\epsilon=0.08$, $\eta=0.5$, $p=3$ and $\gamma=1$}
    \label{fig:enter-label}
\end{figure}
\section{Dissipative case}
\label{xsec4}
We now repeat the previous analysis for the dissipative case, i.e., $\gamma < 1$. As observed in the $V_{rms}$ curves in Section II, the velocities now reach a plateau for a sufficiently large number of collisions. A sufficiently low initial velocity, $V_0=0.2$, was chosen to exhibit both transitions: from the initial plateau at $V_0$ to the growth regime at $n_1$ and from the growth regime to the saturation regime at $n_2$. Fig. 6 presents the behaviors of $V_{rms}$, $V_{max}$, $V_{min}$, $\overline{V}_{max}$ and $\overline{V}_{min}$ for $p=3$, $\epsilon=0.1$, $\eta=0.2$, and $\gamma=0.999$, ensuring that we are on the verge of the transition between the conservative and dissipative dynamics.
\begin{figure}[H]
  \centering
    \centerline{\includegraphics[width=0.85\linewidth]{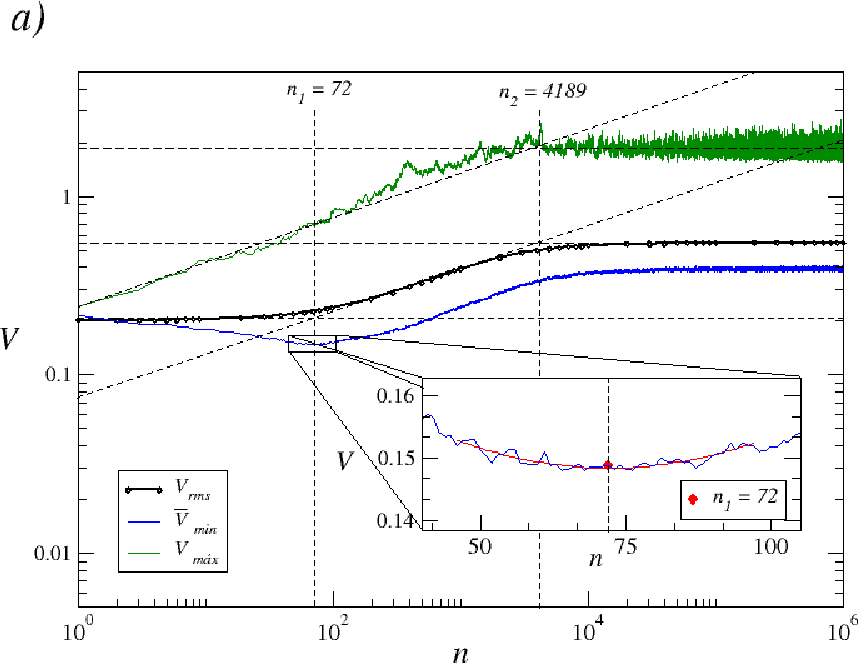}}
    \centerline{\includegraphics[width=0.85\linewidth]{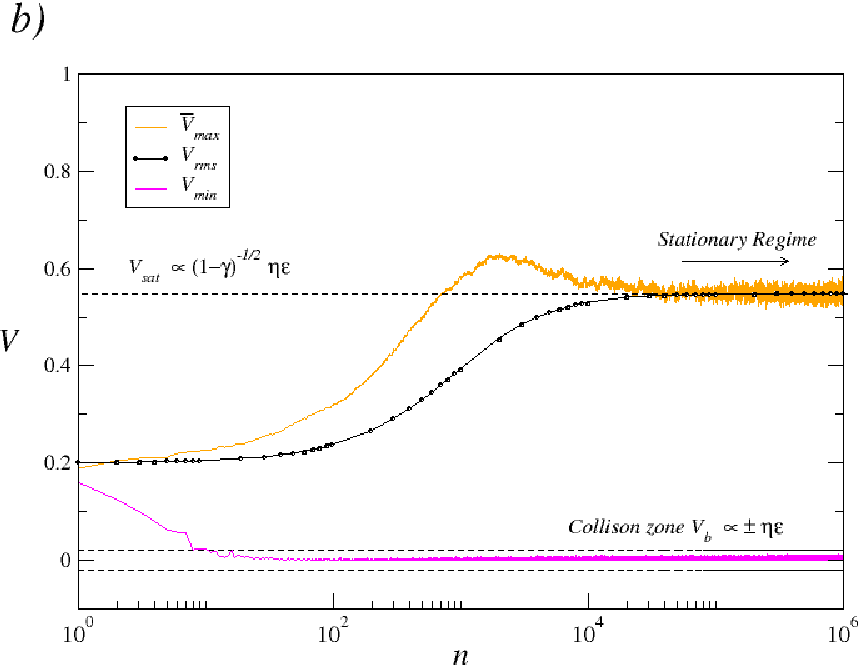}}    
  \caption{(Color online) a) $V_{rms}$, $V_{max}$ and $\overline{V}_{min}$ vs. $n$ for $V_0=0.2$, $\epsilon=0.1$, $\eta=0.2$ and $\gamma=0.999$, with a magnified view of the crossover iteration $n_1$ obtained from the quadratic fit (in red) of the blue curve. As shown, the crossover iterations $n_1$ and $n_2$ were also obtained from the black and green curves. b) $V_{rms}$,$\overline{V}_{max}$ and $V_{min}$ for the same parameters, with $V_{sat}$ and the collision zone at $V_b$ indicated by dotted lines.} \label{}
\end{figure}
Figure 6 a) shows the crossover iterations $n_1$ and $n_2$ obtained by two different methods: (i) from the intersection between the growth and saturation regimes of the curves $V_{max}$ and $V_{rms}$, represented by the dotted lines (ii) from the minimum of the $\overline{V}_{min}$ through a quadratic fit, shown in red. The results found show complete compatibility.

As for Fig, 6 b), the curves of $V_{rms}$, $\overline{V}_{max}$ and $V_{min}$ are shown, where the latter presents the same behavior observed in Fig.4, a decay until the lower limit region, given by the boundary velocity. From this point on, $V_{min}$ oscillates within the collision region described by $V_b \propto \pm \eta \epsilon$. The velocity $\overline{V}_{max}$ presents a growth pattern until the moment in which it crosses the line given by the saturation velocity. This region marks, physically, velocities large enough that the losses due to dissipation, even if minimal, are comparable to the velocities themselves. The saturation velocity $V_{sat}$ is then reached, after a sufficiently large number of collisions, not only for $V_{rms}$ but also for $\overline{V}_{max}$. Such $V_{sat} \propto (1-\gamma)^{-1/2}\eta\epsilon$ acts as an upper limit for the system's velocity.
\begin{figure*}[htb]
    \centering
    \includegraphics[width=0.7\textwidth]{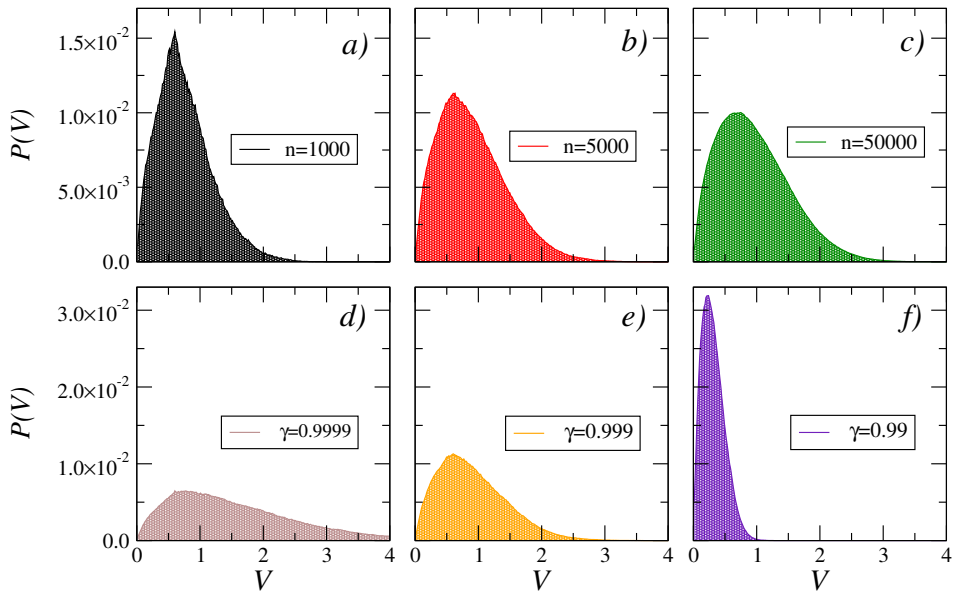}
    \caption{(Color online) Probability density for a set of 3500 particles after for $V_0=0.6$, $\epsilon=0.08$, $\eta=0.5$, $p=3$. Figures a), b), and c) show the behavior for $\gamma=0.999$ and $n=1000$, $n=5000$, and $n=50000$ collisions. Figures d), e) and f), for $n=5000$ collisions, show the behavior for $\gamma=0.99$, $\gamma=0.999$ and $\gamma=0.9999$}
    \label{fig:enter-label}
\end{figure*}
This behavior is corroborated by the distributions presented in Fig. 7, constructed for $V_0=0.6$, $\epsilon=0.08$, $\eta=0.5$, and $p=3$. The figures are robust to the increase in collisions up to $n=50,000$ and to the variation of $\gamma$ near the transition. For a sufficiently small number of collisions, the behavior observed in Fig. 5 is recovered; the distribution is centered symmetrically around the initial velocity $V_0$. The increase in collisions leads to the asymmetric behavior observed in items $a)-c)$, an upper limit inhibits the curve from spreading indefinitely, reaching a steady state, as indicated in Fig. 6b). The behavior is also consistent with the variation of the restitution coefficient $\gamma$, which controls how fast such a distribution flattens. Fig. 7 b) shows, for the same $n=5,000$, a distribution that progressively spreads out as $\gamma$ increases, approaching the transition at $\gamma=1$. It is essential to mention that this upper limit given by the saturation velocity $V_{sat}$ is also proportional to $\gamma$, allowing curves $d)-f)$ to reach different maximum values.

The same distribution can also be characterized analytically by \cite{eu}
\begin{equation}
P(V, n) = \frac{\tau}{\sqrt{4 \pi D n}} \left[ e^{ \left( \frac{-(V - V_0)^2}{4 D n} \right)} - e^{\left( \frac{-(V + V_0)^2}{4 D n} \right)} \right],
\end{equation}
where \( \tau = \text{erf}\left( \frac{V_0}{\sqrt{4 D n}} \right) \) is a normalization constant. This distribution retrieves the behavior \( P(V, 0) = \delta(V - V_0) \) and is constructed using the images method, assuming that there is a "barrier" at $+\infty$ and an absorbing or reflecting one at $0$. This equation only works as expected when the system is sufficiently close to the transition, i.e. $\gamma \rightarrow 1$ since its obtaining arises from the solution of the diffusion equation $\frac{\partial P(V, n)}{\partial n} = D \frac{\partial^2 P(V, n)}{\partial V^2}$.  In the conservative case, we would have to resort to a construction where such an upper limit does not exist, with only one lower "barrier".  

The symmetry breaking that occurs at the transition point $\gamma=1$ is then characterized by the change from a regime where there is only a lower limit for the velocities given by the boundary, in the conservative case, to a second regime, dissipative, where the same lower limit exists alongside an upper limit given by the saturation velocity. This behavior can be verified both in the curves $V_{rms}$, $V_{max}$, $V_{min}$, $\overline{V}_{max}$ and $\overline{V}_{min}$ and in the distribution $P(V)$. A straightforward comparison between Fig. 4 and Fig. 6 shows the same behavior for $V{min}$ and $\overline{V}_{min}$, representing the lower limit, and significantly different behaviors for $V_{max}$ and $\overline{V}_{max}$. After an initial growth, the latter reaches a steady state for a sufficiently large number of collisions due to the dissipation effect, suppressing the FA.

\section{Summary and conclusions}
\label{xsec5}

We characterized the symmetry breaking from a dissipative to a non-dissipative dynamics for a time-dependent billiard. The transition leads to an upper limit alongside the preexisting lower limit defined by the velocity of the boundary. The discussion enabled us to identify the role of the velocity distribution in such symmetry breaking, with the quantity $\overline{V}_{min}$ marking the crossover iteration as the point where it stops decreasing and starts increasing alongside the $\overline{V}_{max}$. This behavior leads to the power law growth regime for $V_{rms}$ and is exhibited in the conservative and dissipative scenarios. The quantities $V_{max}$ and $\overline{V}_{max}$, on the other hand, exhibit significant change after the transition, reaching an upper limit and eventual saturation for the dissipative case, suppressing the FA. The distributions $P(V)$, phenomenological and analytical, corroborate this analysis. Characterizing the symmetry-breaking mechanisms allows a better understanding of the continuous phase transitions exhibited for these systems, marked by scaling invariance, with scaling laws and critical exponents defining the criticality near the transition. Other billiards framed by the LRA conjecture, such as the elliptical and annular \cite{oliveira2012scaling,PhysRevLett.84.867}, that exhibit FA for the conservative case, which can then be suppressed by dissipation, may also be framed by this characterization.

\section*{Conflict of Interest}
The authors declare no conflict of interest.

\section*{Author Contributions}
All authors have contributed equally to the work.
\section*{Funding}
A.K.P.F. acknowledges CAPES for financial support under Grant No. $88887.990665/2024-00$.  E.D.L. acknowledges support from Brazilian agencies CNPq (No. $301318/2019-0, 304398/2023-3)$ and FAPESP (No. $2019/14038-6$ and No. $2021/09519-5)$.

\bibliography{IJAPM}

\begin{thebibliography}{35}%
\makeatletter
\providecommand \@ifxundefined [1]{%
 \@ifx{#1\undefined}
}%
\providecommand \@ifnum [1]{%
 \ifnum #1\expandafter \@firstoftwo
 \else \expandafter \@secondoftwo
 \fi
}%
\providecommand \@ifx [1]{%
 \ifx #1\expandafter \@firstoftwo
 \else \expandafter \@secondoftwo
 \fi
}%
\providecommand \natexlab [1]{#1}%
\providecommand \enquote  [1]{``#1''}%
\providecommand \bibnamefont  [1]{#1}%
\providecommand \bibfnamefont [1]{#1}%
\providecommand \citenamefont [1]{#1}%
\providecommand \href@noop [0]{\@secondoftwo}%
\providecommand \href [0]{\begingroup \@sanitize@url \@href}%
\providecommand \@href[1]{\@@startlink{#1}\@@href}%
\providecommand \@@href[1]{\endgroup#1\@@endlink}%
\providecommand \@sanitize@url [0]{\catcode `\\12\catcode `\$12\catcode
  `\&12\catcode `\#12\catcode `\^12\catcode `\_12\catcode `\%12\relax}%
\providecommand \@@startlink[1]{}%
\providecommand \@@endlink[0]{}%
\providecommand \url  [0]{\begingroup\@sanitize@url \@url }%
\providecommand \@url [1]{\endgroup\@href {#1}{\urlprefix }}%
\providecommand \urlprefix  [0]{URL }%
\providecommand \Eprint [0]{\href }%
\providecommand \doibase [0]{https://doi.org/}%
\providecommand \selectlanguage [0]{\@gobble}%
\providecommand \bibinfo  [0]{\@secondoftwo}%
\providecommand \bibfield  [0]{\@secondoftwo}%
\providecommand \translation [1]{[#1]}%
\providecommand \BibitemOpen [0]{}%
\providecommand \bibitemStop [0]{}%
\providecommand \bibitemNoStop [0]{.\EOS\space}%
\providecommand \EOS [0]{\spacefactor3000\relax}%
\providecommand \BibitemShut  [1]{\csname bibitem#1\endcsname}%
\let\auto@bib@innerbib\@empty
\bibitem [{\citenamefont {Fröhlich}(2024)}]{phase1}%
  \BibitemOpen
  \bibfield  {author} {\bibinfo {author} {\bibfnamefont {J.}~\bibnamefont
  {Fröhlich}},\ }\bibfield  {title} {\bibinfo {title} {Phase transitions,
  spontaneous symmetry breaking, and goldstone's theorem},\ }in\ \href
  {https://doi.org/https://doi.org/10.1016/B978-0-323-90800-9.00275-4} {\emph
  {\bibinfo {booktitle} {Encyclopedia of Condensed Matter Physics (Second
  Edition)}}},\ \bibinfo {editor} {edited by\ \bibinfo {editor} {\bibfnamefont
  {T.}~\bibnamefont {Chakraborty}}}\ (\bibinfo  {publisher} {Academic Press},\
  \bibinfo {address} {Oxford},\ \bibinfo {year} {2024})\ \bibinfo {edition}
  {second edition}\ ed.,\ pp.\ \bibinfo {pages} {158--173}\BibitemShut
  {NoStop}%
\bibitem [{\citenamefont {Chaikin}(1995)}]{principles}%
  \BibitemOpen
  \bibfield  {author} {\bibinfo {author} {\bibfnamefont {L.~T.~C.}\
  \bibnamefont {Chaikin}, \bibfnamefont {P~M}},\ }\href@noop {} {\emph
  {\bibinfo {title} {Principles of Condensed Matter Physics}}}\ (\bibinfo
  {publisher} {Cambridge University Press},\ \bibinfo {year}
  {1995})\BibitemShut {NoStop}%
\bibitem [{\citenamefont {Larsen}\ and\ \citenamefont
  {Nielsen}(2024)}]{manybody1}%
  \BibitemOpen
  \bibfield  {author} {\bibinfo {author} {\bibfnamefont {P.~G.}\ \bibnamefont
  {Larsen}}\ and\ \bibinfo {author} {\bibfnamefont {A.~E.~B.}\ \bibnamefont
  {Nielsen}},\ }\bibfield  {title} {\bibinfo {title} {Phase transitions in
  quantum many-body scars},\ }\href
  {https://doi.org/10.1103/PhysRevResearch.6.L042007} {\bibfield  {journal}
  {\bibinfo  {journal} {Phys. Rev. Res.}\ }\textbf {\bibinfo {volume} {6}},\
  \bibinfo {pages} {L042007} (\bibinfo {year} {2024})}\BibitemShut {NoStop}%
\bibitem [{\citenamefont {Adachi}\ \emph {et~al.}(2022)\citenamefont {Adachi},
  \citenamefont {Takasan},\ and\ \citenamefont {Kawaguchi}}]{manybody2}%
  \BibitemOpen
  \bibfield  {author} {\bibinfo {author} {\bibfnamefont {K.}~\bibnamefont
  {Adachi}}, \bibinfo {author} {\bibfnamefont {K.}~\bibnamefont {Takasan}},\
  and\ \bibinfo {author} {\bibfnamefont {K.}~\bibnamefont {Kawaguchi}},\
  }\bibfield  {title} {\bibinfo {title} {Activity-induced phase transition in a
  quantum many-body system},\ }\href
  {https://doi.org/10.1103/PhysRevResearch.4.013194} {\bibfield  {journal}
  {\bibinfo  {journal} {Phys. Rev. Res.}\ }\textbf {\bibinfo {volume} {4}},\
  \bibinfo {pages} {013194} (\bibinfo {year} {2022})}\BibitemShut {NoStop}%
\bibitem [{\citenamefont {Sadhasivam}\ \emph {et~al.}(2024)\citenamefont
  {Sadhasivam}, \citenamefont {Suzuki}, \citenamefont {Yan},\ and\
  \citenamefont {Sinitsyn}}]{Sadhasivam2024}%
  \BibitemOpen
  \bibfield  {author} {\bibinfo {author} {\bibfnamefont {V.~G.}\ \bibnamefont
  {Sadhasivam}}, \bibinfo {author} {\bibfnamefont {F.}~\bibnamefont {Suzuki}},
  \bibinfo {author} {\bibfnamefont {B.}~\bibnamefont {Yan}},\ and\ \bibinfo
  {author} {\bibfnamefont {N.~A.}\ \bibnamefont {Sinitsyn}},\ }\bibfield
  {title} {\bibinfo {title} {Parametric tuning of quantum phase transitions in
  ultracold reactions},\ }\href {https://doi.org/10.1038/s41467-024-54489-3}
  {\bibfield  {journal} {\bibinfo  {journal} {Nature Communications}\ }\textbf
  {\bibinfo {volume} {15}},\ \bibinfo {pages} {10246} (\bibinfo {year}
  {2024})}\BibitemShut {NoStop}%
\bibitem [{\citenamefont {Aarts}\ \emph {et~al.}(2023)\citenamefont {Aarts},
  \citenamefont {Aichelin}, \citenamefont {Allton}, \citenamefont
  {Athenodorou}, \citenamefont {Bachtis}, \citenamefont {Bonanno},
  \citenamefont {Brambilla}, \citenamefont {Bratkovskaya}, \citenamefont
  {Bruno}, \citenamefont {Caselle}, \citenamefont {Conti}, \citenamefont
  {Contino}, \citenamefont {Cosmai}, \citenamefont {Cuteri}, \citenamefont
  {Del~Debbio}, \citenamefont {D'Elia}, \citenamefont {Dimopoulos},
  \citenamefont {Di~Renzo}, \citenamefont {Galatyuk}, \citenamefont {Guenther},
  \citenamefont {Houtz}, \citenamefont {Karsch}, \citenamefont {Kotov},
  \citenamefont {Lombardo}, \citenamefont {Lucini}, \citenamefont {Maio},
  \citenamefont {Panero}, \citenamefont {Pawlowski}, \citenamefont
  {Pelissetto}, \citenamefont {Philipsen}, \citenamefont {Rago}, \citenamefont
  {Ratti}, \citenamefont {Ryan}, \citenamefont {Sannino}, \citenamefont
  {Sasaki}, \citenamefont {Schicho}, \citenamefont {Schmidt}, \citenamefont
  {Sharma}, \citenamefont {Soloveva}, \citenamefont {Sorba},\ and\
  \citenamefont {Wiese}}]{Aarts2023}%
  \BibitemOpen
  \bibfield  {author} {\bibinfo {author} {\bibfnamefont {G.}~\bibnamefont
  {Aarts}}, \bibinfo {author} {\bibfnamefont {J.}~\bibnamefont {Aichelin}},
  \bibinfo {author} {\bibfnamefont {C.}~\bibnamefont {Allton}}, \bibinfo
  {author} {\bibfnamefont {A.}~\bibnamefont {Athenodorou}}, \bibinfo {author}
  {\bibfnamefont {D.}~\bibnamefont {Bachtis}}, \bibinfo {author} {\bibfnamefont
  {C.}~\bibnamefont {Bonanno}}, \bibinfo {author} {\bibfnamefont
  {N.}~\bibnamefont {Brambilla}}, \bibinfo {author} {\bibfnamefont
  {E.}~\bibnamefont {Bratkovskaya}}, \bibinfo {author} {\bibfnamefont
  {M.}~\bibnamefont {Bruno}}, \bibinfo {author} {\bibfnamefont
  {M.}~\bibnamefont {Caselle}}, \bibinfo {author} {\bibfnamefont
  {C.}~\bibnamefont {Conti}}, \bibinfo {author} {\bibfnamefont
  {R.}~\bibnamefont {Contino}}, \bibinfo {author} {\bibfnamefont
  {L.}~\bibnamefont {Cosmai}}, \bibinfo {author} {\bibfnamefont
  {F.}~\bibnamefont {Cuteri}}, \bibinfo {author} {\bibfnamefont
  {L.}~\bibnamefont {Del~Debbio}}, \bibinfo {author} {\bibfnamefont
  {M.}~\bibnamefont {D'Elia}}, \bibinfo {author} {\bibfnamefont
  {P.}~\bibnamefont {Dimopoulos}}, \bibinfo {author} {\bibfnamefont
  {F.}~\bibnamefont {Di~Renzo}}, \bibinfo {author} {\bibfnamefont
  {T.}~\bibnamefont {Galatyuk}}, \bibinfo {author} {\bibfnamefont {J.~N.}\
  \bibnamefont {Guenther}}, \bibinfo {author} {\bibfnamefont {R.}~\bibnamefont
  {Houtz}}, \bibinfo {author} {\bibfnamefont {F.}~\bibnamefont {Karsch}},
  \bibinfo {author} {\bibfnamefont {A.~Y.}\ \bibnamefont {Kotov}}, \bibinfo
  {author} {\bibfnamefont {M.~P.}\ \bibnamefont {Lombardo}}, \bibinfo {author}
  {\bibfnamefont {B.}~\bibnamefont {Lucini}}, \bibinfo {author} {\bibfnamefont
  {L.}~\bibnamefont {Maio}}, \bibinfo {author} {\bibfnamefont {M.}~\bibnamefont
  {Panero}}, \bibinfo {author} {\bibfnamefont {J.~M.}\ \bibnamefont
  {Pawlowski}}, \bibinfo {author} {\bibfnamefont {A.}~\bibnamefont
  {Pelissetto}}, \bibinfo {author} {\bibfnamefont {O.}~\bibnamefont
  {Philipsen}}, \bibinfo {author} {\bibfnamefont {A.}~\bibnamefont {Rago}},
  \bibinfo {author} {\bibfnamefont {C.}~\bibnamefont {Ratti}}, \bibinfo
  {author} {\bibfnamefont {S.~M.}\ \bibnamefont {Ryan}}, \bibinfo {author}
  {\bibfnamefont {F.}~\bibnamefont {Sannino}}, \bibinfo {author} {\bibfnamefont
  {C.}~\bibnamefont {Sasaki}}, \bibinfo {author} {\bibfnamefont
  {P.}~\bibnamefont {Schicho}}, \bibinfo {author} {\bibfnamefont
  {C.}~\bibnamefont {Schmidt}}, \bibinfo {author} {\bibfnamefont
  {S.}~\bibnamefont {Sharma}}, \bibinfo {author} {\bibfnamefont
  {O.}~\bibnamefont {Soloveva}}, \bibinfo {author} {\bibfnamefont
  {M.}~\bibnamefont {Sorba}},\ and\ \bibinfo {author} {\bibfnamefont {U.-J.}\
  \bibnamefont {Wiese}},\ }\bibfield  {title} {\bibinfo {title} {Phase
  transitions in particle physics: Results and perspectives from lattice
  quantum chromo-dynamics},\ }\href
  {https://doi.org/10.1016/j.ppnp.2023.104070} {\bibfield  {journal} {\bibinfo
  {journal} {Progress in Particle and Nuclear Physics}\ }\textbf {\bibinfo
  {volume} {133}},\ \bibinfo {pages} {104070} (\bibinfo {year}
  {2023})}\BibitemShut {NoStop}%
\bibitem [{\citenamefont {Raz}\ and\ \citenamefont {Levine}(2023)}]{info1}%
  \BibitemOpen
  \bibfield  {author} {\bibinfo {author} {\bibfnamefont {T.}~\bibnamefont
  {Raz}}\ and\ \bibinfo {author} {\bibfnamefont {R.~D.}\ \bibnamefont
  {Levine}},\ }\bibfield  {title} {\bibinfo {title} {The essence of phase
  transitions in condensed matter by an information theoretic approach},\
  }\href {https://doi.org/10.1073/pnas.2310281120} {\bibfield  {journal}
  {\bibinfo  {journal} {Proceedings of the National Academy of Sciences}\
  }\textbf {\bibinfo {volume} {120}},\ \bibinfo {pages} {e2310281120} (\bibinfo
  {year} {2023})}\BibitemShut {NoStop}%
\bibitem [{\citenamefont {Reeves}\ and\ \citenamefont {Pfister}(2021)}]{info2}%
  \BibitemOpen
  \bibfield  {author} {\bibinfo {author} {\bibfnamefont {G.}~\bibnamefont
  {Reeves}}\ and\ \bibinfo {author} {\bibfnamefont {H.~D.}\ \bibnamefont
  {Pfister}},\ }\bibinfo {title} {Understanding phase transitions via mutual
  information and mmse},\ in\ \href@noop {} {\emph {\bibinfo {booktitle}
  {Information-Theoretic Methods in Data Science}}},\ \bibinfo {editor} {edited
  by\ \bibinfo {editor} {\bibfnamefont {M.~R.~D.}\ \bibnamefont {Rodrigues}}\
  and\ \bibinfo {editor} {\bibfnamefont {Y.~C.}\ \bibnamefont {Eldar}}}\
  (\bibinfo  {publisher} {Cambridge University Press},\ \bibinfo {year}
  {2021})\ p.\ \bibinfo {pages} {197–228}\BibitemShut {NoStop}%
\bibitem [{\citenamefont {Leonel}\ \emph {et~al.}(2015)\citenamefont {Leonel},
  \citenamefont {Penalva}, \citenamefont {Teixeira}, \citenamefont
  {Costa~Filho}, \citenamefont {Silva},\ and\ \citenamefont
  {De~Oliveira}}]{leonel2015dynamical}%
  \BibitemOpen
  \bibfield  {author} {\bibinfo {author} {\bibfnamefont {E.~D.}\ \bibnamefont
  {Leonel}}, \bibinfo {author} {\bibfnamefont {J.}~\bibnamefont {Penalva}},
  \bibinfo {author} {\bibfnamefont {R.~M.}\ \bibnamefont {Teixeira}}, \bibinfo
  {author} {\bibfnamefont {R.~N.}\ \bibnamefont {Costa~Filho}}, \bibinfo
  {author} {\bibfnamefont {M.~R.}\ \bibnamefont {Silva}},\ and\ \bibinfo
  {author} {\bibfnamefont {J.~A.}\ \bibnamefont {De~Oliveira}},\ }\bibfield
  {title} {\bibinfo {title} {A dynamical phase transition for a family of
  hamiltonian mappings: A phenomenological investigation to obtain the critical
  exponents},\ }\href@noop {} {\bibfield  {journal} {\bibinfo  {journal}
  {Physics Letters A}\ }\textbf {\bibinfo {volume} {379}},\ \bibinfo {pages}
  {1808} (\bibinfo {year} {2015})}\BibitemShut {NoStop}%
\bibitem [{\citenamefont {Oliveira}\ and\ \citenamefont
  {Leonel}(2013)}]{oliveira2013some}%
  \BibitemOpen
  \bibfield  {author} {\bibinfo {author} {\bibfnamefont {D.~F.}\ \bibnamefont
  {Oliveira}}\ and\ \bibinfo {author} {\bibfnamefont {E.~D.}\ \bibnamefont
  {Leonel}},\ }\bibfield  {title} {\bibinfo {title} {Some dynamical properties
  of a classical dissipative bouncing ball model with two nonlinearities},\
  }\href@noop {} {\bibfield  {journal} {\bibinfo  {journal} {Physica A:
  Statistical Mechanics and its Applications}\ }\textbf {\bibinfo {volume}
  {392}},\ \bibinfo {pages} {1762} (\bibinfo {year} {2013})}\BibitemShut
  {NoStop}%
\bibitem [{\citenamefont {Sethna}(2021)}]{sethna2021statistical}%
  \BibitemOpen
  \bibfield  {author} {\bibinfo {author} {\bibfnamefont {J.~P.}\ \bibnamefont
  {Sethna}},\ }\href@noop {} {\emph {\bibinfo {title} {Statistical mechanics:
  entropy, order parameters, and complexity}}},\ Vol.~\bibinfo {volume} {14}\
  (\bibinfo  {publisher} {Oxford University Press, USA},\ \bibinfo {year}
  {2021})\BibitemShut {NoStop}%
\bibitem [{\citenamefont {Landau}(1969)}]{landau}%
  \BibitemOpen
  \bibfield  {author} {\bibinfo {author} {\bibfnamefont {L.}~\bibnamefont
  {Landau}},\ }\href@noop {} {\emph {\bibinfo {title} {Statistical physics}}},\
  Vol.~\bibinfo {volume} {6}\ (\bibinfo  {publisher} {Pergamon},\ \bibinfo
  {year} {1969})\BibitemShut {NoStop}%
\bibitem [{\citenamefont {Ehrenfest}(1933)}]{ehrenfest1933phase}%
  \BibitemOpen
  \bibfield  {author} {\bibinfo {author} {\bibfnamefont {P.}~\bibnamefont
  {Ehrenfest}},\ }\bibfield  {title} {\bibinfo {title} {Phase transitions in
  the usual and generalized sense, classified according to the singularities of
  the thermodynamic potential},\ }\href@noop {} {\bibfield  {journal} {\bibinfo
   {journal} {Proceedings of the Amsterdam Academy}\ }\textbf {\bibinfo
  {volume} {36}},\ \bibinfo {pages} {153–157} (\bibinfo {year}
  {1933})}\BibitemShut {NoStop}%
\bibitem [{\citenamefont {Reif}(2009)}]{reif2009fundamentals}%
  \BibitemOpen
  \bibfield  {author} {\bibinfo {author} {\bibfnamefont {F.}~\bibnamefont
  {Reif}},\ }\href@noop {} {\emph {\bibinfo {title} {Fundamentals of
  statistical and thermal physics}}}\ (\bibinfo  {publisher} {Waveland Press},\
  \bibinfo {year} {2009})\BibitemShut {NoStop}%
\bibitem [{\citenamefont {Leonel}\ \emph {et~al.}(2020)\citenamefont {Leonel},
  \citenamefont {Yoshida},\ and\ \citenamefont
  {de~Oliveira}}]{leonel2020characterization}%
  \BibitemOpen
  \bibfield  {author} {\bibinfo {author} {\bibfnamefont {E.~D.}\ \bibnamefont
  {Leonel}}, \bibinfo {author} {\bibfnamefont {M.}~\bibnamefont {Yoshida}},\
  and\ \bibinfo {author} {\bibfnamefont {J.~A.}\ \bibnamefont {de~Oliveira}},\
  }\bibfield  {title} {\bibinfo {title} {Characterization of a continuous phase
  transition in a chaotic system},\ }\href@noop {} {\bibfield  {journal}
  {\bibinfo  {journal} {Europhysics Letters}\ }\textbf {\bibinfo {volume}
  {131}},\ \bibinfo {pages} {20002} (\bibinfo {year} {2020})}\BibitemShut
  {NoStop}%
\bibitem [{\citenamefont {Chernov}\ and\ \citenamefont
  {Markarian}(2006)}]{chernov2006chaotic}%
  \BibitemOpen
  \bibfield  {author} {\bibinfo {author} {\bibfnamefont {N.}~\bibnamefont
  {Chernov}}\ and\ \bibinfo {author} {\bibfnamefont {R.}~\bibnamefont
  {Markarian}},\ }\href@noop {} {\emph {\bibinfo {title} {Chaotic
  billiards}}},\ \bibinfo {number} {127}\ (\bibinfo  {publisher} {American
  Mathematical Soc.},\ \bibinfo {year} {2006})\BibitemShut {NoStop}%
\bibitem [{\citenamefont {Berry}(1981)}]{berry1981regularity}%
  \BibitemOpen
  \bibfield  {author} {\bibinfo {author} {\bibfnamefont {M.~V.}\ \bibnamefont
  {Berry}},\ }\bibfield  {title} {\bibinfo {title} {Regularity and chaos in
  classical mechanics, illustrated by three deformations of a
  circular'billiard'},\ }\href@noop {} {\bibfield  {journal} {\bibinfo
  {journal} {European Journal of Physics}\ }\textbf {\bibinfo {volume} {2}},\
  \bibinfo {pages} {91} (\bibinfo {year} {1981})}\BibitemShut {NoStop}%
\bibitem [{\citenamefont {Bunimovich}\ and\ \citenamefont
  {Dettmann}(2005)}]{bunimovich2005open}%
  \BibitemOpen
  \bibfield  {author} {\bibinfo {author} {\bibfnamefont {L.}~\bibnamefont
  {Bunimovich}}\ and\ \bibinfo {author} {\bibfnamefont {C.}~\bibnamefont
  {Dettmann}},\ }\bibfield  {title} {\bibinfo {title} {Open circular billiards
  and the riemann hypothesis},\ }\href@noop {} {\bibfield  {journal} {\bibinfo
  {journal} {Physical review letters}\ }\textbf {\bibinfo {volume} {94}},\
  \bibinfo {pages} {100201} (\bibinfo {year} {2005})}\BibitemShut {NoStop}%
\bibitem [{\citenamefont {Dettmann}\ and\ \citenamefont
  {Georgiou}(2009)}]{dettmann2009survival}%
  \BibitemOpen
  \bibfield  {author} {\bibinfo {author} {\bibfnamefont {C.~P.}\ \bibnamefont
  {Dettmann}}\ and\ \bibinfo {author} {\bibfnamefont {O.}~\bibnamefont
  {Georgiou}},\ }\bibfield  {title} {\bibinfo {title} {Survival probability for
  the stadium billiard},\ }\href@noop {} {\bibfield  {journal} {\bibinfo
  {journal} {Physica D: Nonlinear Phenomena}\ }\textbf {\bibinfo {volume}
  {238}},\ \bibinfo {pages} {2395} (\bibinfo {year} {2009})}\BibitemShut
  {NoStop}%
\bibitem [{\citenamefont {Lozej}\ and\ \citenamefont
  {Robnik}(2018)}]{lozej2018aspects}%
  \BibitemOpen
  \bibfield  {author} {\bibinfo {author} {\bibfnamefont {{\v{C}}.}~\bibnamefont
  {Lozej}}\ and\ \bibinfo {author} {\bibfnamefont {M.}~\bibnamefont {Robnik}},\
  }\bibfield  {title} {\bibinfo {title} {Aspects of diffusion in the stadium
  billiard},\ }\href@noop {} {\bibfield  {journal} {\bibinfo  {journal}
  {Physical Review E}\ }\textbf {\bibinfo {volume} {97}},\ \bibinfo {pages}
  {012206} (\bibinfo {year} {2018})}\BibitemShut {NoStop}%
\bibitem [{\citenamefont {Lopac}\ \emph {et~al.}(2002)\citenamefont {Lopac},
  \citenamefont {Mrkonji{\'c}},\ and\ \citenamefont
  {Radi{\'c}}}]{lopac2002chaotic}%
  \BibitemOpen
  \bibfield  {author} {\bibinfo {author} {\bibfnamefont {V.}~\bibnamefont
  {Lopac}}, \bibinfo {author} {\bibfnamefont {I.}~\bibnamefont
  {Mrkonji{\'c}}},\ and\ \bibinfo {author} {\bibfnamefont {D.}~\bibnamefont
  {Radi{\'c}}},\ }\bibfield  {title} {\bibinfo {title} {Chaotic dynamics and
  orbit stability in the parabolic oval billiard},\ }\href@noop {} {\bibfield
  {journal} {\bibinfo  {journal} {Physical Review E}\ }\textbf {\bibinfo
  {volume} {66}},\ \bibinfo {pages} {036202} (\bibinfo {year}
  {2002})}\BibitemShut {NoStop}%
\bibitem [{\citenamefont {Lichtenberg}\ and\ \citenamefont
  {Lieberman}(2013)}]{lichtenberg2013regular}%
  \BibitemOpen
  \bibfield  {author} {\bibinfo {author} {\bibfnamefont {A.~J.}\ \bibnamefont
  {Lichtenberg}}\ and\ \bibinfo {author} {\bibfnamefont {M.~A.}\ \bibnamefont
  {Lieberman}},\ }\href@noop {} {\emph {\bibinfo {title} {Regular and chaotic
  dynamics}}},\ Vol.~\bibinfo {volume} {38}\ (\bibinfo  {publisher} {Springer
  Science \& Business Media},\ \bibinfo {year} {2013})\BibitemShut {NoStop}%
\bibitem [{\citenamefont {Dembowski}\ \emph {et~al.}(2000)\citenamefont
  {Dembowski}, \citenamefont {Gr\"af}, \citenamefont {Heine}, \citenamefont
  {Hofferbert}, \citenamefont {Rehfeld},\ and\ \citenamefont
  {Richter}}]{PhysRevLett.84.867}%
  \BibitemOpen
  \bibfield  {author} {\bibinfo {author} {\bibfnamefont {C.}~\bibnamefont
  {Dembowski}}, \bibinfo {author} {\bibfnamefont {H.-D.}\ \bibnamefont
  {Gr\"af}}, \bibinfo {author} {\bibfnamefont {A.}~\bibnamefont {Heine}},
  \bibinfo {author} {\bibfnamefont {R.}~\bibnamefont {Hofferbert}}, \bibinfo
  {author} {\bibfnamefont {H.}~\bibnamefont {Rehfeld}},\ and\ \bibinfo {author}
  {\bibfnamefont {A.}~\bibnamefont {Richter}},\ }\bibfield  {title} {\bibinfo
  {title} {First experimental evidence for chaos-assisted tunneling in a
  microwave annular billiard},\ }\href
  {https://doi.org/10.1103/PhysRevLett.84.867} {\bibfield  {journal} {\bibinfo
  {journal} {Phys. Rev. Lett.}\ }\textbf {\bibinfo {volume} {84}},\ \bibinfo
  {pages} {867} (\bibinfo {year} {2000})}\BibitemShut {NoStop}%
\bibitem [{\citenamefont {Wang}\ \emph {et~al.}(2021)\citenamefont {Wang},
  \citenamefont {Liu}, \citenamefont {Liu}, \citenamefont {Xiao}, \citenamefont
  {Wang}, \citenamefont {Fan}, \citenamefont {Han}, \citenamefont {Ge},\ and\
  \citenamefont {Song}}]{Wang2021}%
  \BibitemOpen
  \bibfield  {author} {\bibinfo {author} {\bibfnamefont {S.}~\bibnamefont
  {Wang}}, \bibinfo {author} {\bibfnamefont {S.}~\bibnamefont {Liu}}, \bibinfo
  {author} {\bibfnamefont {Y.}~\bibnamefont {Liu}}, \bibinfo {author}
  {\bibfnamefont {S.}~\bibnamefont {Xiao}}, \bibinfo {author} {\bibfnamefont
  {Z.}~\bibnamefont {Wang}}, \bibinfo {author} {\bibfnamefont {Y.}~\bibnamefont
  {Fan}}, \bibinfo {author} {\bibfnamefont {J.}~\bibnamefont {Han}}, \bibinfo
  {author} {\bibfnamefont {L.}~\bibnamefont {Ge}},\ and\ \bibinfo {author}
  {\bibfnamefont {Q.}~\bibnamefont {Song}},\ }\bibfield  {title} {\bibinfo
  {title} {Direct observation of chaotic resonances in optical microcavities},\
  }\href {https://doi.org/10.1038/s41377-021-00578-7} {\bibfield  {journal}
  {\bibinfo  {journal} {Light: Science \& Applications}\ }\textbf {\bibinfo
  {volume} {10}},\ \bibinfo {pages} {135} (\bibinfo {year} {2021})}\BibitemShut
  {NoStop}%
\bibitem [{\citenamefont {Berglund}\ and\ \citenamefont
  {Kunz}(1996)}]{Berglund1996}%
  \BibitemOpen
  \bibfield  {author} {\bibinfo {author} {\bibfnamefont {N.}~\bibnamefont
  {Berglund}}\ and\ \bibinfo {author} {\bibfnamefont {H.}~\bibnamefont
  {Kunz}},\ }\bibfield  {title} {\bibinfo {title} {Integrability and ergodicity
  of classical billiards in a magnetic field},\ }\href
  {https://doi.org/10.1007/BF02183708} {\bibfield  {journal} {\bibinfo
  {journal} {Journal of Statistical Physics}\ }\textbf {\bibinfo {volume}
  {83}},\ \bibinfo {pages} {81} (\bibinfo {year} {1996})}\BibitemShut {NoStop}%
\bibitem [{\citenamefont {Chandramouli}\ \emph {et~al.}(2020)\citenamefont
  {Chandramouli}, \citenamefont {Srivastav},\ and\ \citenamefont
  {Kumar}}]{Electronictransport}%
  \BibitemOpen
  \bibfield  {author} {\bibinfo {author} {\bibfnamefont {R.~S.}\ \bibnamefont
  {Chandramouli}}, \bibinfo {author} {\bibfnamefont {R.~K.}\ \bibnamefont
  {Srivastav}},\ and\ \bibinfo {author} {\bibfnamefont {S.}~\bibnamefont
  {Kumar}},\ }\bibfield  {title} {\bibinfo {title} {Electronic transport in
  chaotic mesoscopic cavities: A kwant and random matrix theory based
  exploration},\ }\href {https://doi.org/10.1063/5.0026039} {\bibfield
  {journal} {\bibinfo  {journal} {Chaos: An Interdisciplinary Journal of
  Nonlinear Science}\ }\textbf {\bibinfo {volume} {30}},\ \bibinfo {pages}
  {123120} (\bibinfo {year} {2020})},\ \Eprint
  {https://arxiv.org/abs/https://pubs.aip.org/aip/cha/article-pdf/doi/10.1063/5.0026039/14109797/123120\_1\_online.pdf}
  {https://pubs.aip.org/aip/cha/article-pdf/doi/10.1063/5.0026039/14109797/123120\_1\_online.pdf}
  \BibitemShut {NoStop}%
\bibitem [{\citenamefont {Exner}\ \emph {et~al.}(2006)\citenamefont {Exner},
  \citenamefont {Hejčík},\ and\ \citenamefont {Šeba}}]{EXNER2006445}%
  \BibitemOpen
  \bibfield  {author} {\bibinfo {author} {\bibfnamefont {P.}~\bibnamefont
  {Exner}}, \bibinfo {author} {\bibfnamefont {P.}~\bibnamefont {Hejčík}},\
  and\ \bibinfo {author} {\bibfnamefont {P.}~\bibnamefont {Šeba}},\ }\bibfield
   {title} {\bibinfo {title} {Approximations by graphs and emergence of global
  structures},\ }\href
  {https://doi.org/https://doi.org/10.1016/S0034-4877(06)80031-5} {\bibfield
  {journal} {\bibinfo  {journal} {Reports on Mathematical Physics}\ }\textbf
  {\bibinfo {volume} {57}},\ \bibinfo {pages} {445} (\bibinfo {year}
  {2006})}\BibitemShut {NoStop}%
\bibitem [{\citenamefont {Miao}\ \emph {et~al.}(2007)\citenamefont {Miao},
  \citenamefont {Wijeratne}, \citenamefont {Zhang}, \citenamefont {Coskun},
  \citenamefont {Bao},\ and\ \citenamefont {Lau}}]{science.1144359}%
  \BibitemOpen
  \bibfield  {author} {\bibinfo {author} {\bibfnamefont {F.}~\bibnamefont
  {Miao}}, \bibinfo {author} {\bibfnamefont {S.}~\bibnamefont {Wijeratne}},
  \bibinfo {author} {\bibfnamefont {Y.}~\bibnamefont {Zhang}}, \bibinfo
  {author} {\bibfnamefont {U.~C.}\ \bibnamefont {Coskun}}, \bibinfo {author}
  {\bibfnamefont {W.}~\bibnamefont {Bao}},\ and\ \bibinfo {author}
  {\bibfnamefont {C.~N.}\ \bibnamefont {Lau}},\ }\bibfield  {title} {\bibinfo
  {title} {Phase-coherent transport in graphene quantum billiards},\ }\href
  {https://doi.org/10.1126/science.1144359} {\bibfield  {journal} {\bibinfo
  {journal} {Science}\ }\textbf {\bibinfo {volume} {317}},\ \bibinfo {pages}
  {1530} (\bibinfo {year} {2007})},\ \Eprint
  {https://arxiv.org/abs/https://www.science.org/doi/pdf/10.1126/science.1144359}
  {https://www.science.org/doi/pdf/10.1126/science.1144359} \BibitemShut
  {NoStop}%
\bibitem [{\citenamefont {Loskutov}\ \emph {et~al.}(1999)\citenamefont
  {Loskutov}, \citenamefont {Ryabov},\ and\ \citenamefont
  {Akinshin}}]{loskutov1999mechanism}%
  \BibitemOpen
  \bibfield  {author} {\bibinfo {author} {\bibfnamefont {A.~Y.}\ \bibnamefont
  {Loskutov}}, \bibinfo {author} {\bibfnamefont {A.}~\bibnamefont {Ryabov}},\
  and\ \bibinfo {author} {\bibfnamefont {L.}~\bibnamefont {Akinshin}},\
  }\bibfield  {title} {\bibinfo {title} {Mechanism of fermi acceleration in
  dispersing billiards with time-dependent boundaries},\ }\href@noop {}
  {\bibfield  {journal} {\bibinfo  {journal} {Journal of Experimental and
  Theoretical Physics}\ }\textbf {\bibinfo {volume} {89}},\ \bibinfo {pages}
  {966} (\bibinfo {year} {1999})}\BibitemShut {NoStop}%
\bibitem [{\citenamefont {Loskutov}\ \emph {et~al.}(2000)\citenamefont
  {Loskutov}, \citenamefont {Ryabov},\ and\ \citenamefont
  {Akinshin}}]{loskutov2000properties}%
  \BibitemOpen
  \bibfield  {author} {\bibinfo {author} {\bibfnamefont {A.}~\bibnamefont
  {Loskutov}}, \bibinfo {author} {\bibfnamefont {A.}~\bibnamefont {Ryabov}},\
  and\ \bibinfo {author} {\bibfnamefont {L.}~\bibnamefont {Akinshin}},\
  }\bibfield  {title} {\bibinfo {title} {Properties of some chaotic billiards
  with time-dependent boundaries},\ }\href@noop {} {\bibfield  {journal}
  {\bibinfo  {journal} {Journal of Physics A: Mathematical and General}\
  }\textbf {\bibinfo {volume} {33}},\ \bibinfo {pages} {7973} (\bibinfo {year}
  {2000})}\BibitemShut {NoStop}%
\bibitem [{\citenamefont {Oliveira}\ and\ \citenamefont
  {Leonel}(2010)}]{oliveira2010suppressing}%
  \BibitemOpen
  \bibfield  {author} {\bibinfo {author} {\bibfnamefont {D.~F.}\ \bibnamefont
  {Oliveira}}\ and\ \bibinfo {author} {\bibfnamefont {E.~D.}\ \bibnamefont
  {Leonel}},\ }\bibfield  {title} {\bibinfo {title} {Suppressing fermi
  acceleration in a two-dimensional non-integrable time-dependent oval-shaped
  billiard with inelastic collisions},\ }\href@noop {} {\bibfield  {journal}
  {\bibinfo  {journal} {Physica A: Statistical Mechanics and its Applications}\
  }\textbf {\bibinfo {volume} {389}},\ \bibinfo {pages} {1009} (\bibinfo {year}
  {2010})}\BibitemShut {NoStop}%
\bibitem [{\citenamefont {Leonel}\ \emph {et~al.}(2016)\citenamefont {Leonel},
  \citenamefont {Galia}, \citenamefont {Barreiro},\ and\ \citenamefont
  {Oliveira}}]{leonel2016thermodynamics}%
  \BibitemOpen
  \bibfield  {author} {\bibinfo {author} {\bibfnamefont {E.~D.}\ \bibnamefont
  {Leonel}}, \bibinfo {author} {\bibfnamefont {M.~V.~C.}\ \bibnamefont
  {Galia}}, \bibinfo {author} {\bibfnamefont {L.~A.}\ \bibnamefont
  {Barreiro}},\ and\ \bibinfo {author} {\bibfnamefont {D.~F.}\ \bibnamefont
  {Oliveira}},\ }\bibfield  {title} {\bibinfo {title} {Thermodynamics of a
  time-dependent and dissipative oval billiard: A heat transfer and billiard
  approach},\ }\href@noop {} {\bibfield  {journal} {\bibinfo  {journal}
  {Physical Review E}\ }\textbf {\bibinfo {volume} {94}},\ \bibinfo {pages}
  {062211} (\bibinfo {year} {2016})}\BibitemShut {NoStop}%
\bibitem [{\citenamefont {Hansen}\ \emph {et~al.}(2018)\citenamefont {Hansen},
  \citenamefont {Ciro}, \citenamefont {Caldas},\ and\ \citenamefont
  {Leonel}}]{hansen}%
  \BibitemOpen
  \bibfield  {author} {\bibinfo {author} {\bibfnamefont {M.}~\bibnamefont
  {Hansen}}, \bibinfo {author} {\bibfnamefont {D.}~\bibnamefont {Ciro}},
  \bibinfo {author} {\bibfnamefont {I.~L.}\ \bibnamefont {Caldas}},\ and\
  \bibinfo {author} {\bibfnamefont {E.~D.}\ \bibnamefont {Leonel}},\ }\bibfield
   {title} {\bibinfo {title} {Explaining a changeover from normal to super
  diffusion in time-dependent billiards},\ }\href@noop {} {\bibfield  {journal}
  {\bibinfo  {journal} {Europhysics Letters.}\ }\textbf {\bibinfo {volume}
  {121}} (\bibinfo {year} {2018})}\BibitemShut {NoStop}%
\bibitem [{\citenamefont {da~Fonseca}\ \emph {et~al.}(2025)\citenamefont
  {da~Fonseca}, \citenamefont {Silveira}, \citenamefont {Kuwana}, \citenamefont
  {Oliveira}, ,\ and\ \citenamefont {Leonel}}]{eu}%
  \BibitemOpen
  \bibfield  {author} {\bibinfo {author} {\bibfnamefont {A.~K.~P.}\
  \bibnamefont {da~Fonseca}}, \bibinfo {author} {\bibfnamefont {F.~A.~O.}\
  \bibnamefont {Silveira}}, \bibinfo {author} {\bibfnamefont {C.~M.}\
  \bibnamefont {Kuwana}}, \bibinfo {author} {\bibfnamefont {D.~F.~M.}\
  \bibnamefont {Oliveira}}, ,\ and\ \bibinfo {author} {\bibfnamefont {E.~D.}\
  \bibnamefont {Leonel}},\ }\bibfield  {title} {\bibinfo {title} {Transition
  from bounded to unbounded energy in a time-dependent billiard},\ }\href@noop
  {} {\bibfield  {journal} {\bibinfo  {journal} {Physical Review E}\ }
  (\bibinfo {year} {2025})}\BibitemShut {NoStop}%
\bibitem [{\citenamefont {Oliveira}\ and\ \citenamefont
  {Robnik}(2012)}]{oliveira2012scaling}%
  \BibitemOpen
  \bibfield  {author} {\bibinfo {author} {\bibfnamefont {D.~F.}\ \bibnamefont
  {Oliveira}}\ and\ \bibinfo {author} {\bibfnamefont {M.}~\bibnamefont
  {Robnik}},\ }\bibfield  {title} {\bibinfo {title} {Scaling invariance in a
  time-dependent elliptical billiard},\ }\href@noop {} {\bibfield  {journal}
  {\bibinfo  {journal} {International Journal of Bifurcation and Chaos}\
  }\textbf {\bibinfo {volume} {22}},\ \bibinfo {pages} {1250207} (\bibinfo
  {year} {2012})}\BibitemShut {NoStop}%
\end{thebibliography}%

\end{document}